\documentclass[twocolumn,twocolappendix,iop,apj]{emulateapj}
\usepackage{times}
\usepackage{comment}
\usepackage[normalem]{ulem}
\usepackage[usenames, dvipsnames]{color}

\newcommand{\mynote}[1]{#1}

\graphicspath{{images/}}

\newcommand{\degree}{\ensuremath{^\circ }}

\newcommand{\fermilat}{\emph{Fermi}--LAT}

\newcommand{\galprop}{{\sc GALPROP}}

\newcommand{\helmod}{\textsc{HelMod}}

\slugcomment{Draft: \today}

\shorttitle{Local interstellar spectrum of cosmic-ray electrons}
\shortauthors{Boschini~et~al.}

\usepackage{amsmath} 

\begin{document}


\title{
\helmod{} in the works: from direct observations to the local interstellar spectrum\\ of cosmic-ray electrons 
}


\author{
M.~J.~Boschini\altaffilmark{1,2},  S.~{Della~Torre}\altaffilmark{1}, M.~Gervasi\altaffilmark{1,3}, D.~Grandi\altaffilmark{1},
G.~J\'{o}hannesson\altaffilmark{4,5}, 
G.~{La~Vacca}\altaffilmark{1}, N.~Masi\altaffilmark{6},
I.~V.~Moskalenko\altaffilmark{7,8}, 
S.~Pensotti\altaffilmark{1,3}, T.~A.~Porter\altaffilmark{7,8}, L.~Quadrani\altaffilmark{6,9}, P.~G.~Rancoita\altaffilmark{1},
D.~Rozza\altaffilmark{1,3}, and M.~Tacconi\altaffilmark{1,3}
}


\altaffiltext{1}{INFN, Milano-Bicocca, Milano, Italy}
\altaffiltext{2}{also CINECA, Segrate, Milano, Italy}
\altaffiltext{3}{also Physics Department, University of Milano-Bicocca, Milano, Italy}
\altaffiltext{4}{Science Institute, University of Iceland, Dunhaga 3, IS-107 Reykjavik, Iceland}
\altaffiltext{5}{also NORDITA,  Roslagstullsbacken 23, 106 91 Stockholm, Sweden}
\altaffiltext{6}{INFN, Bologna, Italy}
\altaffiltext{7}{Hansen Experimental Physics Laboratory, Stanford University, Stanford, CA 94305}
\altaffiltext{8}{Kavli Institute for Particle Astrophysics and Cosmology, Stanford University, Stanford, CA 94305}
\altaffiltext{9}{also, Physics Department, University of Bologna, Bologna, Italy}


\begin{abstract}

The local interstellar spectrum (LIS) of cosmic-ray (CR) electrons for the energy range 1 MeV to 1 TeV is derived using the most recent experimental results combined with the state-of-the-art models for CR propagation in the Galaxy and in the heliosphere. Two propagation packages, \galprop{} and \helmod{}, are combined to provide a single framework that is run to reproduce direct measurements of CR species at different modulation levels, and at both polarities of the solar magnetic field. An iterative maximum-likelihood method is developed that uses GALPROP-predicted LIS as input to \helmod{}, which provides the modulated spectra for specific time periods of the selected experiments for model-data comparison. The optimized HelMod parameters are then used to adjust \galprop{} parameters to predict a refined LIS with the procedure repeated subject to a convergence criterion. The parameter optimization uses an extensive data set of proton spectra from 1997--2015. The proposed CR electron LIS accommodates both the low-energy interstellar spectra measured by Voyager 1 as well as the high-energy observations by PAMELA and AMS-02 that are made deep in the heliosphere; it also accounts for Ulysses counting rate features measured out of the ecliptic plane. The interstellar and heliospheric propagation parameters derived in this study agree well with our earlier results for CR protons, helium nuclei, and anti-protons propagation and LIS obtained in the same framework.

\end{abstract}


\keywords{cosmic rays --- diffusion --- elementary particles --- interplanetary medium --- ISM: general --- solar system: general}

\section{Introduction} \label{Introduction} \label{Intr}

Electrons in the cosmic radiation were identified for the first time about fifty years after the discovery of CRs \citep{1961PhRvL...6..193M,1961PhRvL...6..125E}. Subsequently, the origin of the observed spectrum of CR electrons has been one of the most important questions in CR physics. Early CR electron measurements of increasing precision and expanding energy range were made over a series of balloon flights by different experiments \citep[e.g.,][]{
1969ApJ...158..771F,
1975ApJ...199..669B,
1976ApJ...204..927H,
1984ApJ...287..622G,
1994ApJ...436..769G,
1995ICRC....3....1B,
1998ApJ...498..779B,
2000ApJ...532..653B,
2001ApJ...559..973T,
2002A&A...392..287G}.   
However, the experimental scatter was large because the CR electron spectrum is steeply falling with increasing energy, and the background of heavier CR species is high.

The first high-statistics measurements of the all-electron CR spectrum over a wide energy range were made by the \emph{Fermi} Large Area Telescope (\fermilat{}) launched in 2008 \citep{2009ApJ...697.1071A}. These measurements showed that the all-electron spectrum is flatter than expected with an index about --3 over the energy range 7--1000 GeV \citep{2009PhRvL.102r1101A,2010PhRvD..82i2004A}. PAMELA data \citep[1--625 GeV,][]{2011PhRvL.106t1101A} generally confirmed the \fermilat{} results albeit with larger error bars, as did the higher precision data from AMS-02 \citep[0.5-1000 GeV,][]{2014PhRvL.113v1102A}. Even though the precise AMS-02 data showed deviations from the earlier \fermilat\ measurements that are significant due to high statistics and consequently very small error bars, the absolute difference is only $\sim$10\% above 20 GeV vs.\ a factor of $\sim$3--4 in the pre-{\it Fermi} era. Note that the latest \fermilat\ all-electron spectrum (7--2000 GeV) obtained using a revised event reconstruction 
and background rejection analysis \citep[][]{2017PhRvD..95h2007A} agrees well with AMS-02 results. Above $\sim$1 TeV, the all-electron spectrum falls rapidly \citep[H.E.S.S.,][]{2008PhRvL.101z1104A,2009A&A...508..561A}. The first ever measurement of the all-electron spectrum for energies $\lesssim100$ MeV outside of the heliosphere has been made by Voyager 1 that reached the heliopause in 2012 \citep{2013Sci...341..150S,2016ApJ...831...18C}.

The strong interest in the CR electron spectrum during the last decade is also fueled by the PAMELA discovery of a continuous rise of the positron fraction up to $\sim100$~GeV \citep{2009Natur.458..607A}, and expectations of spectral features at very-high energies associated with local CR accelerators \citep[e.g.,][]{2004ApJ...601..340K}. The latter are yet to be found, although the dedicated experiment CALET has been operating on the International Space Station (ISS) since August 19, 2015 \citep{2017APh....91....1A}, and the ISS-CREAM, which was launched to the ISS on August 14, 2017, is also deployed there to make CR measurements in the multi-TeV range \citep{2014AdSpR..53.1451S}. 

The discovery of the rise of the positron fraction by PAMELA, contrary to the expectations based on the pure secondary production of positrons in energetic CR interactions with the interstellar gas \citep{1982ApJ...254..391P,1998ApJ...493..694M}, was the first clear evidence of new phenomena detected in CRs, even though the first hints of it appeared in data collected by earlier experiments. The TS93 apparatus launched on a balloon from Fort Sumner, NM, in 1993 measured a flat positron fraction $0.078\pm0.016$ in the range $\sim$5--60 GeV \citep{1996ApJ...457L.103G}. Subsequent balloon-borne flights by CAPRICE94 in 1994 \citep{2000ApJ...532..653B}, the HEAT-$e^\pm$ instrument in 1994 and 1995 \citep{1997ApJ...482L.191B}, and HEAT-pbar instrument in 2000 \citep{2004PhRvL..93x1102B}, indicated that the positron flux did not fall-off as quickly as expected. 
However, the experimental error bars in these early experiments were too large to provide convincing evidence for a new phenomenon.

Following the PAMELA discovery the rise of the positron fraction up to 200~GeV was confirmed by the \fermilat\ \citep{2012PhRvL.108a1103A}, where the geomagnetic field (the ``East-West effect'') was used to provide the charge sign separation, and then up to $\sim500$ GeV with higher precision by AMS-02 \citep{2014PhRvL.113l1101A,2014PhRvL.113v1102A}. These measurements stimulated an extensive discussion of the origin of the rising positron fraction with dozens of different hypotheses proposed in the literature.
They range from conventional astrophysics to non-standard model physics involving various types of dark matter particles. 
A component with similar origin could be also present in the electron spectrum \citep[e.g.,][]{RozzaJHEA_2015}. 

High-precision measurements of both electrons and positrons over a wide energy range are thus of critical importance toward unveiling the origin of the excess positrons. Meanwhile, the $e^\pm$ spectra and the positron fraction below $\sim$10 GeV was found to depend on the solar activity \citep[PAMELA,][]{2016PhRvL.116x1105A}. The determination of the true electron LIS is, therefore, of considerable interest for the astrophysics and particle physics communities. In the present paper, the same method \mynote{-- including the treatment of errors --} is employed as for the recently published studies devoted to the LIS of CR protons, helium nuclei, and anti-protons \citep{2017ApJ...840..115B}. 

\section{\galprop{} and \helmod{} codes}

In this paper, we use a recently developed version of the \helmod\footnote{\mynote{
In this work we use \helmod{} version 3.5, available from http://www.helmod.org/. The origin of the \helmod{} code goes back to the work by \citeauthor*{1999NuPhS..78...26G} (\citeyear{1999NuPhS..78...26G}) \citep[see, for instance,][]{doi:10.1142/9789812702708_0004,DellaTorre2009,Bobik2011ApJ,DellaTorre2012,DellaTorre2013AdvAstro,BobikEtAl2016,2017HelMod}. It has been under continuous development since that time. \label{helmod-link}}} 2D Monte Carlo code for heliospheric CR propagation \citep[][]{Bobik2011ApJ,DellaTorre2013AdvAstro,2017HelMod} combined with the \galprop{}\footnote{http://galprop.stanford.edu} code for interstellar CR propagation \citep{2016ApJ...824...16J,2017ApJ...846...67P} to take advantage of the progress made in the recent CR electron measurements and to derive a self-consistent electron LIS. The \helmod{} code includes all relevant effects and, thus, a full description of the diffusion tensor. \helmod\ enables accurate calculations for the heliospheric modulation effect over arbitrary epochs and is easily interfaced with \galprop{}.

\subsection{Galactic CR propagation with the \galprop{} code}
\label{galprop}

The \galprop\ code has been under development since the mid-90s \citep{1998ApJ...493..694M,1998ApJ...509..212S} and is the {\it de facto} standard code for calculating the propagation of CRs and their associated interstellar emissions.
It solves the CR transport equation for a given source distribution and boundary conditions for all CR species. \galprop\ includes all relevant transport and energy loss/gain processes, such as a galactic wind (advection), diffusive reacceleration in the ISM, energy losses, nuclear fragmentation, radioactive decay, and the production of secondary particles and isotopes. The numerical solution of the transport equation can be obtained using different solvers, including a Crank-Nicholson implicit second-order scheme as well as an explicit method. The spatial boundary conditions assume free particle escape. For a given halo size the diffusion coefficient as a function of momentum is determined by fitting model parameters to CR nuclei secondary-to-primary ratios.  

The \galprop{} code computes a full network of CR primary, secondary and tertiary species from input source abundances. Starting with the heaviest primary nucleus typically considered ($^{64}$Ni, $A=64$) the propagation solution is used to compute the source term for its spallation products $A-1$, $A-2$, and so forth. These are propagated in turn, and so on down in mass to protons, secondary $e^\pm$, and $\bar{p}$. The inelastically scattered $p$ and $\bar{p}$ are treated as separate components (secondary $p$, tertiary $\bar{p}$). \galprop{} includes a description for the processes of K-capture, electron capture by bare CR nuclei and stripping, as well as knock-on electrons. More details are given in \citet{2006ApJ...642..902P}, \citet{2007ARNPS..57..285S}, \citet{2011CoPhC.182.1156V}, and \citet{2016ApJ...824...16J}, as well as the description of the most recent version of \galprop{} (v.~56) -- see \citet{PoS(ICRC2017)279} and \citet{2017ApJ...846...67P}, and references therein.

\subsection{\helmod{} code for heliospheric CR transport}
\label{Sect::Helmod}

\galprop{} provides the predictions for the LIS of all CR species. However, they cannot be compared to the direct CR measurements made at Earth's orbit, or generally in the inner heliosphere, because of the effect of the so-called heliospheric or solar modulation.
This modulation is the combined effect of the expanding magnetic fields and the solar wind (SW) whose properties depend on the level of solar activity~\citep[e.g., see][]{2017ApJ...840..115B,2017HelMod}.

The propagation of CRs in the heliosphere was first studied by \citet{1965P&SS...13....9P}, who formulated the transport equation~\citep[also called the Parker equation -- see, e.g., the discussion in][and reference therein]{Bobik2011ApJ}:

\begin{align}
\label{EQ::FPE}
 \frac{\partial U}{\partial t}= &\frac{\partial}{\partial x_i} \left( K^S_{ij}\frac{\partial \mathrm{U} }{\partial x_j}\right)\\
&+\frac{1}{3}\frac{\partial V_{ \mathrm{sw},i} }{\partial x_i} \frac{\partial }{\partial T}\left(\alpha_{\mathrm{rel} }T\mathrm{U} \right)
- \frac{\partial}{\partial x_i} [ (V_{ \mathrm{sw},i}+v_{d,i})\mathrm{U}],\nonumber
\end{align}
where $U$ is the number density of Galactic CR particles per unit of kinetic energy $T$ (GeV/nucleon), $t$ is time, $V_{ \mathrm{sw},i}$ is the SW velocity along the axis $x_i$, $K^S_{ij}$ is the symmetric part of the diffusion tensor, $v_{d,i}$ is the particle magnetic drift velocity (related to the anti-symmetric part of the diffusion tensor), and $\alpha_{\mathrm{rel} }=\frac{T+2m_r c^2}{T+m_r c^2} $, with $m_r$ -- the particle rest mass per nucleon in units of GeV/nucleon.
The terms in the Parker equation describe:
(i) the \textit{diffusion} of Galactic CRs scattered by magnetic turbulences,
(ii) the \textit{adiabatic} energy losses/gains due to the propagation in the expanding magnetic fields carried in the SW,
(iii) an \textit{effective convection} resulting from the SW convection with velocity $\vec{V}_{\rm sw}$,
and (iv) the \textit{drift effects} related to the drift velocity ($\vec{v}_{\rm drift}$).
Overall, the heliospheric modulation results in energy losses and supression of the fluxes of CR species compared to the LIS that are energy- and charge-sign-dependent. These effects are controlled by the polarity of the solar magnetic field and by the level of solar activity.

The particle transport within the heliosphere, from the Termination Shock (TS) to Earth's orbit, is treated in this paper using the \helmod{} code.
%
%
\helmod{} integrates the \citet{1965P&SS...13....9P} transport equation using a Monte Carlo approach involving stochastic differential equations; for further details of the method and code see \citet{Bobik2011ApJ,BobikEtAl2016}.

\mynote{
In previous models of CR propagation in the heliosphere, the parallel diffusion coefficient ($K_{||}$) was assumed to have
a sharp break at $\sim$1 GV, in the transitional region between the two regimes at high and low rigidities~\citep[e.g., see][]{perko1987,alanko2007,Strauss2011,Bobik2011ApJ}. However, as the accuracy of the collected data increased, it becomes clear that a smooth transition between the two regimes is necessary. The functional form of such a transition that is currently employed in \helmod{} \citep[see Equation 5 in][]{2017HelMod} is consistent with those presented in \citet{BurgerHattingh1998} for the same rigidity interval.}

The normalization of the parallel component $K_{||}$ of the symmetric part of the diffusion tensor $K^S_{ij}$ is determined by the so-called diffusion parameter $K_0$, as defined by Eq.~(2) of \citet[][and references therein]{2017HelMod}. In turn, the diffusion parameter $K_0$ includes a correction factor that rescales the absolute value proportionally to the drift contribution. This correction factor is evaluated in \citet{2017ApJ...840..115B} using the proton spectrum during the period of positive polarity of the heliospheric magnetic field (HMF), and accounts for the presence of the latitudinal structure in the spatial distribution of Galactic CRs. 
The same correction factor is now applied to electron propagation ($q<0$) during the negative HMF polarity period\footnote{A similar correction has to be evaluated for the negative-charge particle diffusion during the positive HMF polarity period ($qA<0$). The negative-charge particles are subject to a correction that is opposite to the one applied to the 
positive-charge particles.} ($A<0$), so that an equivalent scaling\footnote{\helmod{} Parameters -- usually determined at 1 AU -- are used for the properties of any heliospheric sector, according to the time required by the solar wind coming from the Sun to reach such a region \citep{Bobik2011ApJ,2017HelMod}. When this is not accounted for there is an effective time delay in the correlation between time variations of the parameters of the solar magnetic field, as measured at Earth, and the observed intensity variations of GCRs \citep[see, e.g.,][and references therein]{2017TomassettiTimeLag}.} is applied to periods with $qA>0$.

\mynote{The drift treatment in \helmod{} follows the formalism originally developed by \citet{Potgieter85} and refined using Parker's magnetic field with polar correction described in \citet{DellaTorre2013AdvAstro}. During high activity periods the heliospheric magnetic field is far from being considered regular, therefore, we introduced a correction factor suppressing any drift velocity at solar maximum.}

As discussed by \citet{2017HelMod}, the validity of the \helmod{} code is verified down to about 1 GV rigidities (equivalent to $\sim$1 GeV in kinetic energy for electrons). Lower rigidities/energies are not considered in the present work because to do so requires additional refinement for the description of the solar modulation in the outer heliosphere -- between TS and interstellar space \citep[see, e.g.,][]{2011ApJ...735..128S,2017NatAs...1E.115D} -- as well as inclusion of the turbulence in the calculation of the drift coefficient \citep[see, e.g.,][]{2017ApJ...841..107E}. However, Voyager 1 electron data is used as a guideline. 

\begin{deluxetable}{rrlc}
\tablecolumns{4}
\tablewidth{0mm}
\tablecaption{Best-fit propagation parameters for electrons\label{tbl-1}}
\tablehead{
\colhead{N} &
\multicolumn{2}{c}{Parameter\quad} &
\colhead{Best Value}
}
\startdata
1 & $z_h$,& kpc &4.0 \\
2 & $D_{0}$,& $10^{28}$ cm$^{2}$ s$^{-1}$  &4.3 \\
3 & $\delta$ &&0.405 \\
4 & $V_{\rm Alf}$,& km s$^{-1}$ &31 \\
5 & $dV_{\rm conv}/dz$,& km s$^{-1}$ kpc$^{-1}$ & 9.8
\enddata
\end{deluxetable}

\begin{deluxetable}{cc}
\tablecolumns{2}
\tablewidth{0mm}
\tablecaption{Electron injection spectrum
\label{tbl-2}}
\tablehead{
\colhead{Parameters} &
\colhead{Values} 
}
\startdata
$R_{0}$ &190 MV  \\
$R_{1}$ &6 GV  \\
$R_{2}$ &95 GV \\
${\gamma}_0$ &2.57 \\
${\gamma}_1$ &1.40 \\
${\gamma}_2$ &2.80 \\
${\gamma}_3$ &2.40/2.54\tablenotemark{a}
\enddata
\tablenotetext{a}{If an additional component to the electron spectrum is added, see a discussion in Section~\ref{Sect::results}.}
\end{deluxetable}

\begin{figure*}
\begin{center}
\includegraphics[width=0.72\textwidth]{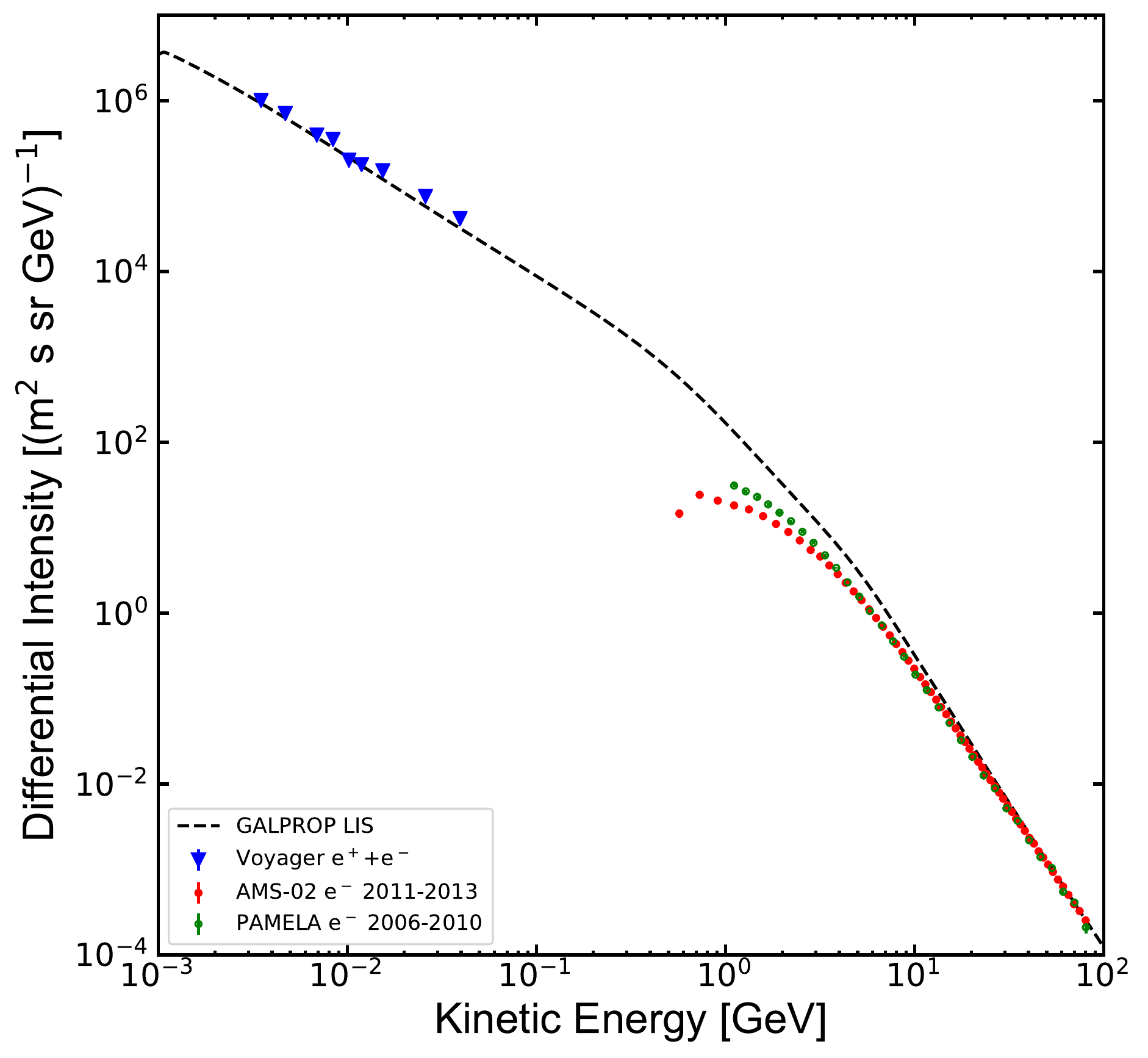}
\caption{The electron LIS (dashed) as derived from the MCMC procedure compared with AMS-02, PAMELA and Voyager 1 measurements (see text).}
 \label{fig:GALPROPLISS}
 \end{center}
\end{figure*}

\section{Interstellar propagation}\label{Sec:IntersProp}

The tuning procedure employed in this paper is the same that was used by \citet{2017ApJ...840..115B}. A short description of the method is provided below.

The Markov Chain Monte Carlo (MCMC) interface to v.56 of \galprop{} was adapted from CosRayMC~\citep{2012PhRvD..85d3507L} and, in general, from the COSMOMC package \citep{2002PhRvD..66j3511L}.
An iterative procedure was developed that calculates LIS with \galprop{}, passing the results to \helmod{} to produce the modulated spectra for specific time periods for comparison with AMS-02 data, which are the observational constraints. The goodness estimator of the parameter scan is the natural logarithm of the likelihood. For computational convenience this is built using $\chi^2$ from all observables: hundreds of thousands of samples were generated and the Log-Likelihood used to accept or reject each sample. The scan is terminated when the Log-Likelihood is maximized.

The basic features of CR propagation in the Galaxy are well-known, but the exact values of propagation parameters depend on the assumed propagation model and accuracy of selected CR data. Therefore, the MCMC procedure is used to determine the propagation parameters employing the best available CR measurements. The five propagation parameters that have the largest effect on the overall shape of CR spectra were left free in the scan that used a 2D \galprop{} model: the Galactic halo half-width $z_h$, the normalization of the diffusion coefficient $D_0$ and the index of its rigidity dependence $\delta$, the Alfv\'en velocity $V_{\rm Alf}$, and the gradient of the convection velocity $dV_{\rm conv}/dz$ ($V_{\rm conv}=0$ in the plane, $z=0$).
The spatial distribution of CRs near the Sun depends only weakly on the chosen radial size of the Galaxy if its distance is farther than the halo size \citep[e.g.,][]{2012ApJ...750....3A}. The radial boundary is therefore set to 20~kpc.

The best values for the main propagation parameters tuned to the AMS-02 data are listed in Table~\ref{tbl-1}. The values are similar to those obtained by \citet{2017ApJ...840..115B}, within the quoted error bands, while the convection velocity $V_{\rm conv}$ is set to 0 in the plane. For example, to get a more consistent electron LIS, the Alfv\'en velocity $V_{\rm Alf}$ was increased by $\sim2$~km~s$^{-1}$. As already discussed by \citet{2017ApJ...840..115B}, simultaneous inclusion of both reacceleration and convection is needed to describe the high precision AMS-02 data, particularly in the range below 20 GV where the modulation effects on CR spectra are significant. For more details the reader is referred to the above-mentioned paper. 


The MCMC procedure is used only for first step to define a consistent set for the Galactic CR propagation parameters. The \helmod\ module was then used for a methodical calibration of the LIS spectral parameters.
Parameters of the injection spectra, such as spectral indices $\gamma_i$ and the break rigidities $R_i$, were left free, but their exact values depend on the solar modulation, so the low-energy parts of the spectra are tuned together with the solar modulation parameters as described below.


To refine the LIS description smoothing features to the breaks in the injection spectrum were added. Reproducing the electron spectrum from MeV to TeV energies requires an injection spectrum with three spectral breaks.
MCMC scans in $\gamma_i$ and $R_i$ were performed using CR electron measurements by AMS-02 \citep{2014PhRvL.113l1102A} and by Voyager~1 \citep{2016ApJ...831...18C} as constraints. At the next step, these parameters were slightly modified together with the solar modulation parameters in order to find the best-fit solution for the electron LIS, as explained by \citet{2017ApJ...840..115B}.
Reproduction of the low-energy electron LIS measurements by Voyager 1 requires a break around $R_{0}\sim190$ MV.
The resulting best-fit spectral parameters are shown in Table~\ref{tbl-2}. 



Note that the only data available to tune the electron LIS below AMS-02 energies are coming from Voyager~1.
Unfortunately, the Electron Telescope (TET) aboard the Voyager~1 spacecraft cannot discriminate between electrons and positrons, so it provides only the all-electron spectrum.
On the other hand, \galprop{} calculations indicate that the secondary positron fraction decreases as energy decreases being $\lesssim35$\% at its maximum contribution for $\sim200$ MeV energies, and becomes as small as a few per cent or less below $\sim$20 MeV \citep[e.g.,][]{2008ApJ...682..400P}.
Therefore, assuming that only electrons are present in CRs at low energies the maximum error in the results at these energies would be $\sim30$\%.

\begin{figure}[tb!]
\centerline{
\includegraphics[width=0.49\textwidth]{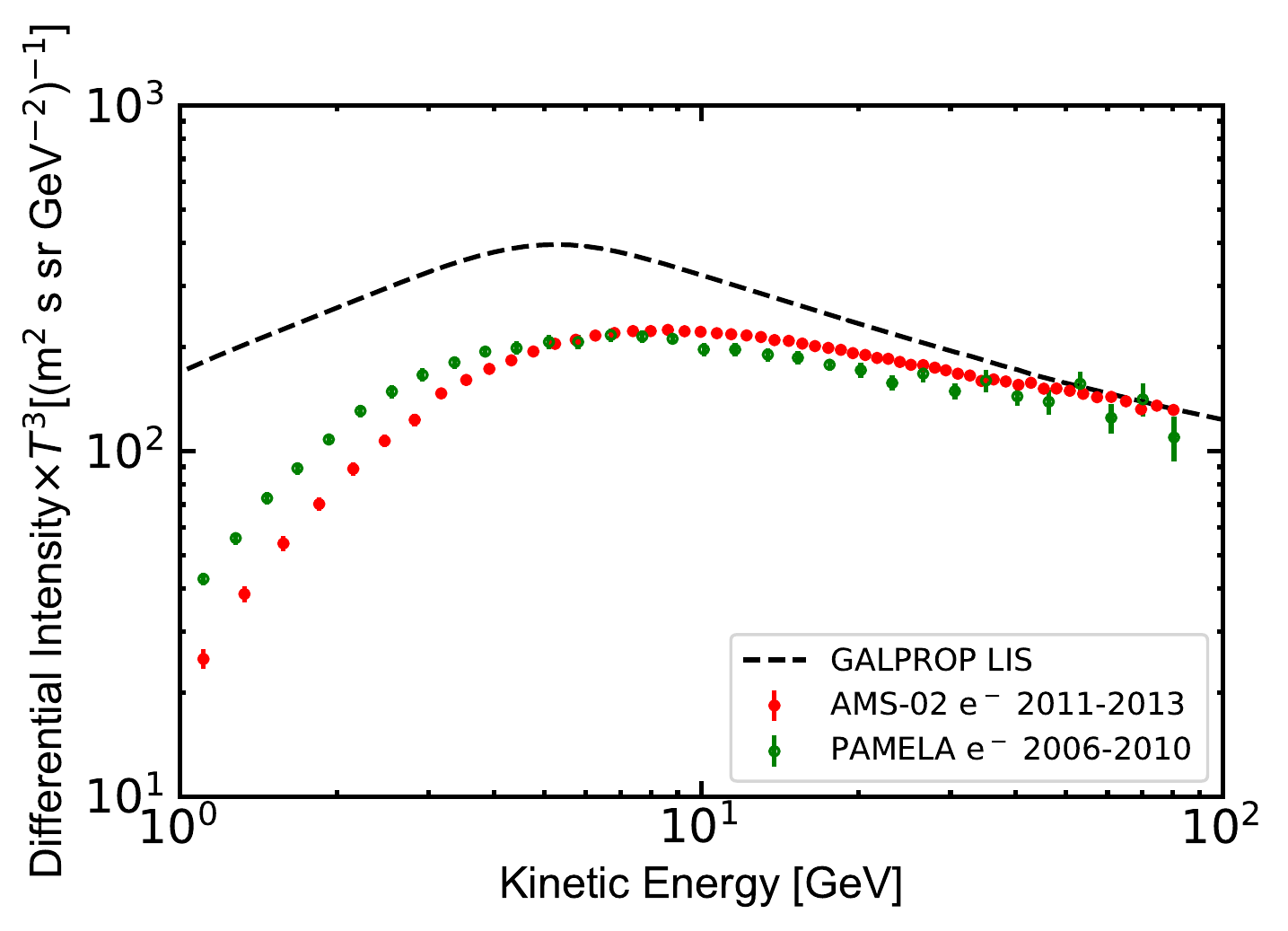}
}
\caption{The proposed electron LIS is compared with 
high-energy all-electron data from AMS-02 and PAMELA experiments.
 }
\label{fig:low_high_LISData}
\end{figure}

\begin{deluxetable*}{crrcc}[tb!]
\tablecolumns{5}
\tablewidth{0mm}
\tablecaption{Normalization corrections applied to the electron LIS\label{tbl-norm}}
\tablehead{
\colhead{Dataset group} &
\colhead{Experiment} &
\colhead{Time span}&
\colhead{Normalization correction}&
\colhead{Reference}
}
\startdata
a)	& PAMELA & 5 years integrated spectrum & 0.81 & \citet{2011PhRvL.106t1101A} \\
b)	& PAMELA & 6 months integrated spectrum & 0.9 & \citet{2015ApJ...810..142A}\\
c)	& AMS-02 & 3 years integrated spectrum & 1.0 & \citet{2014PhRvL.113l1102A}
\enddata
\end{deluxetable*}

\begin{figure*}[tb!]
\centerline{
 \includegraphics[width=0.49\textwidth]{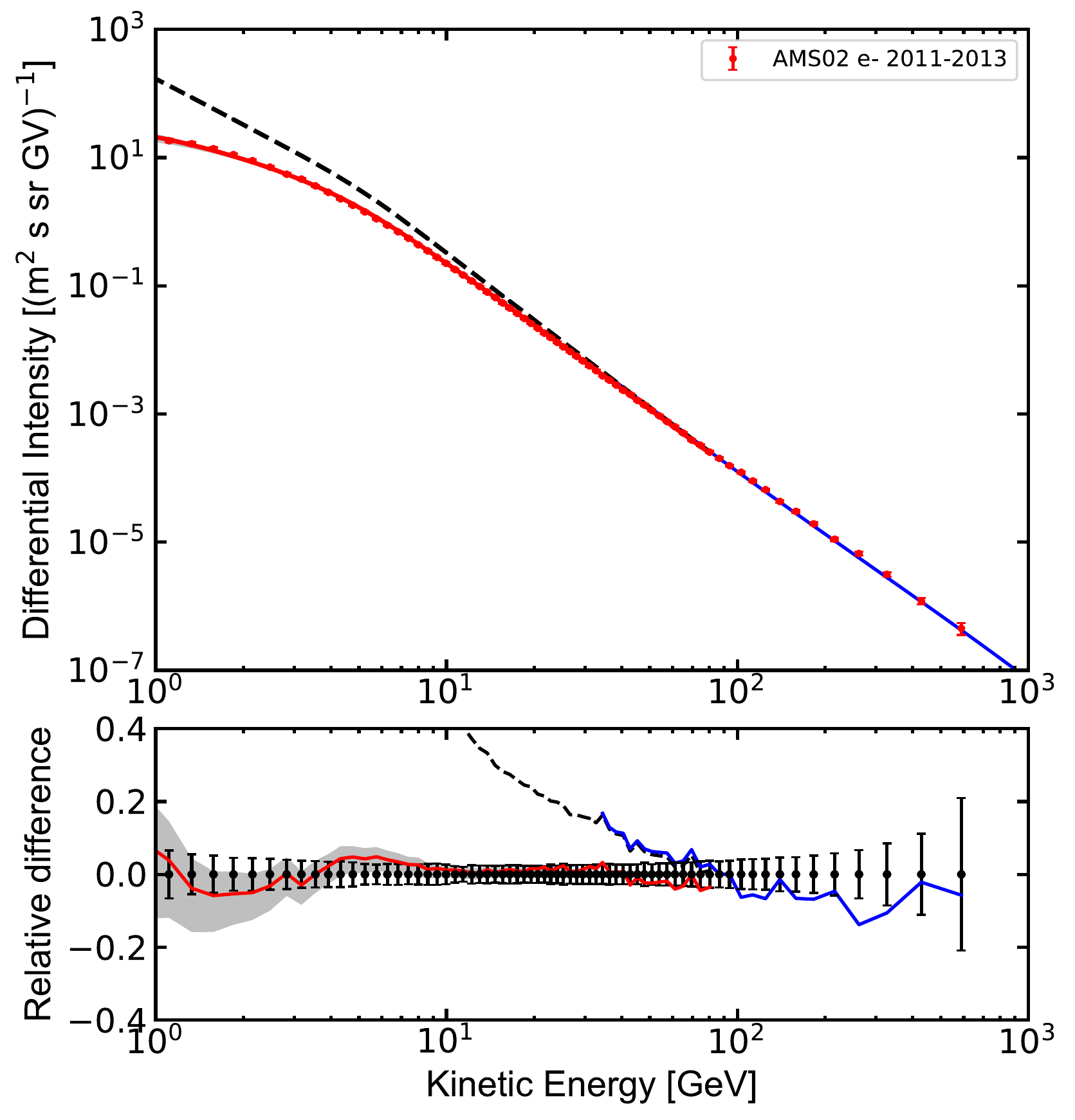}\hfill
 \includegraphics[width=0.49\textwidth]{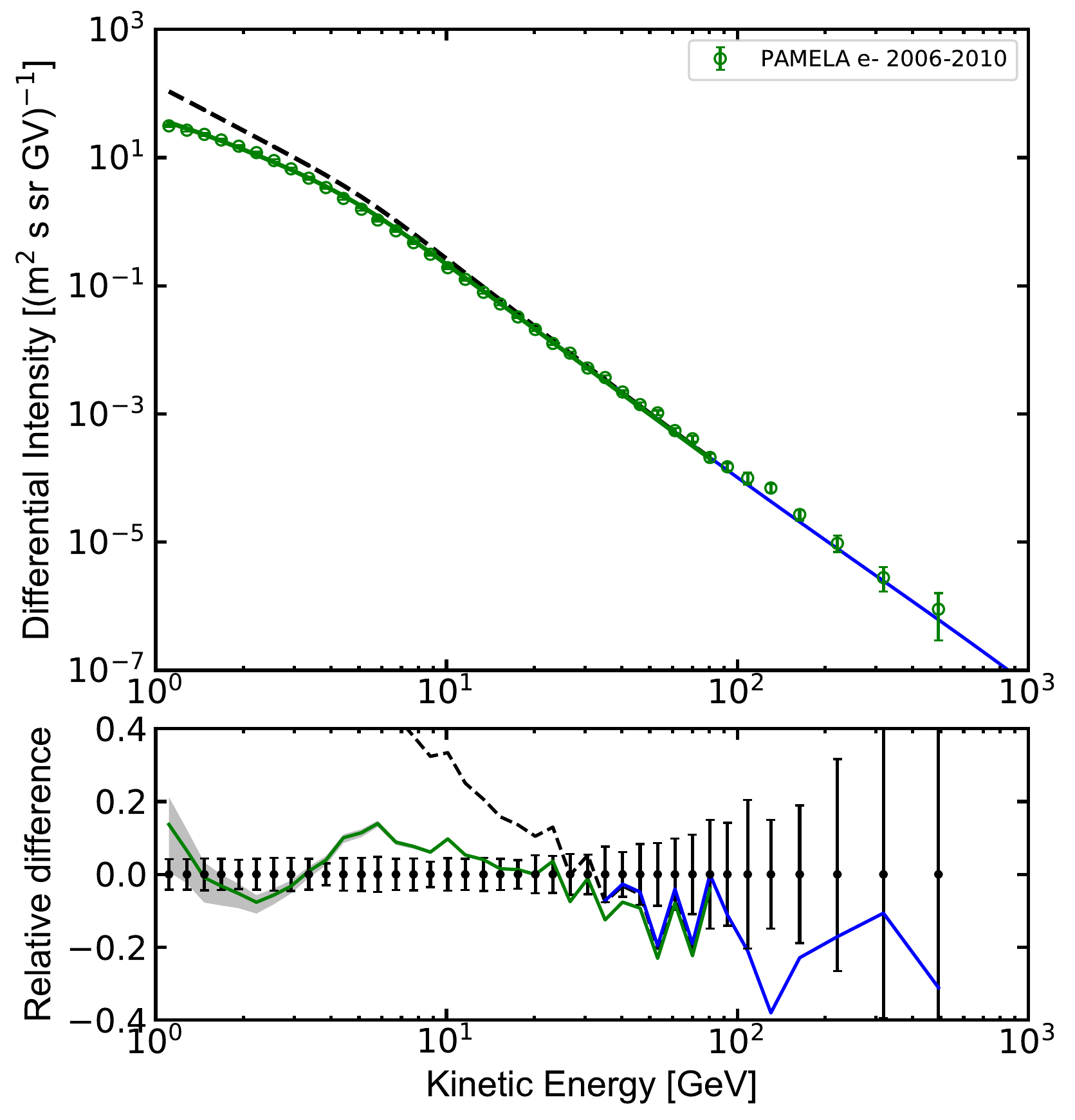}
 }
 \caption{Differential intensity of CR electrons for AMS-02 2011-2013 (left) and PAMELA 2006-2010 (right) datasets. Points represent experimental data, the black dashed line is the \galprop{} LIS, and the red/green solid lines are the computed modulated spectra. The blue solid line represents the expected LIS including the high energy electron excess contribution (see text). The bottom panel shows the relative difference between the numerical solutions and the experimental data.
}
 \label{fig:ElectronAMSPAMELA}
\end{figure*}

\begin{figure*}[tb!]
 \centerline{
 \includegraphics[width=0.49\textwidth]{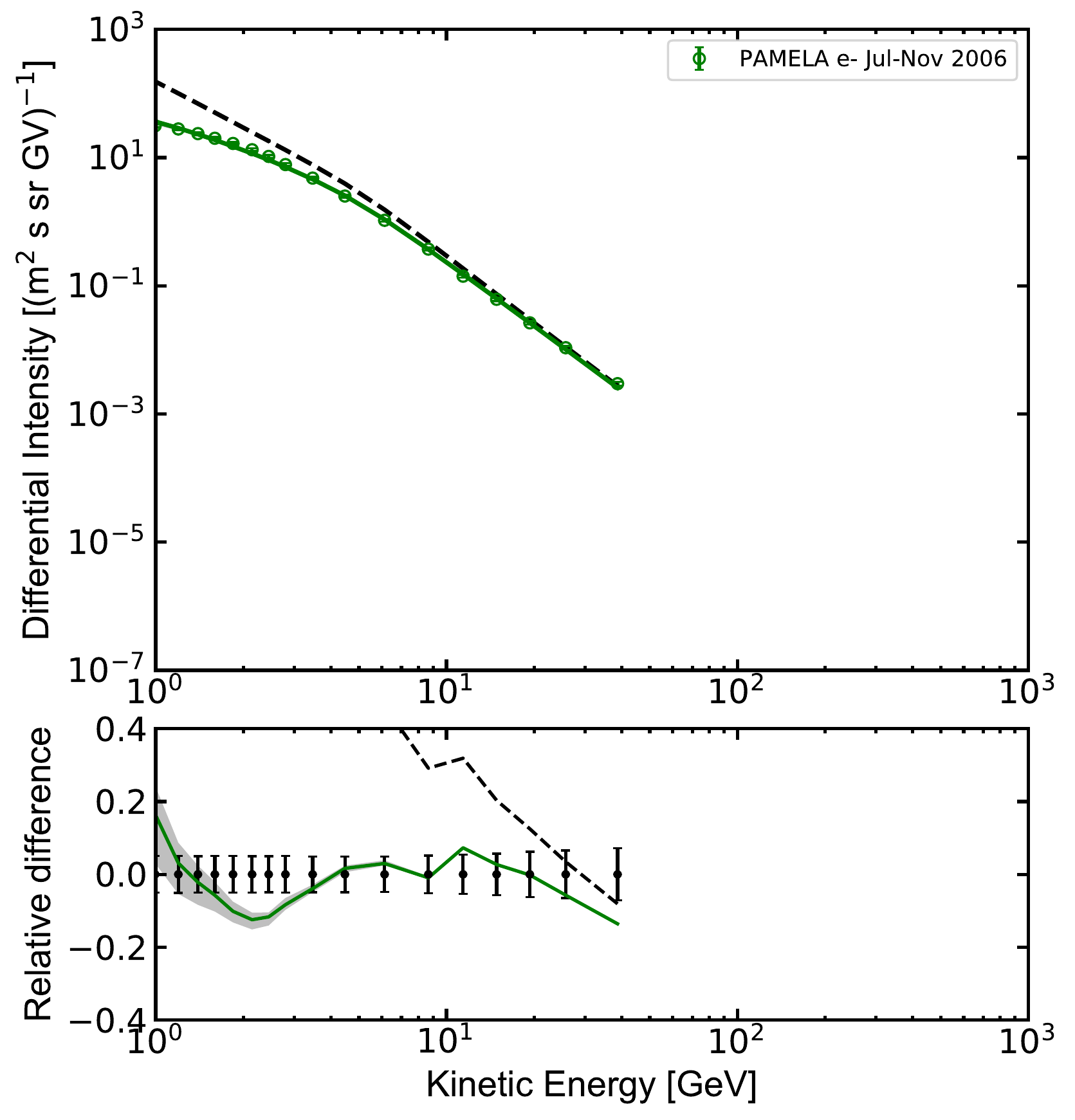}\hfill
 \includegraphics[width=0.49\textwidth]{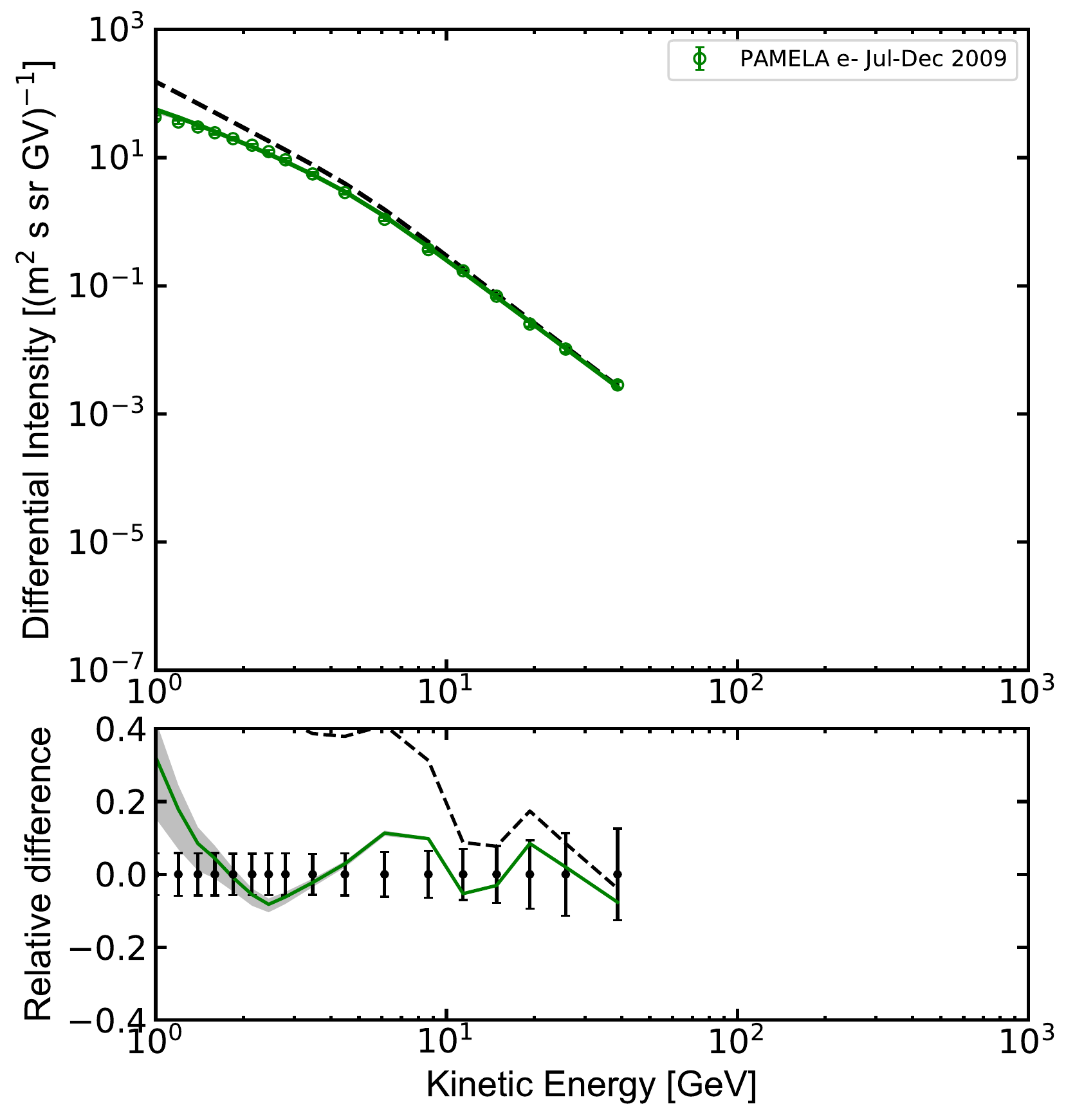}
 }
 \caption{Differential intensity of CR electrons for PAMELA 2006 (left) and PAMELA 2009 (right) datasets. Points represent experimental data, the black dashed line is the \galprop{} LIS, and green solid lines are the computed modulated spectra. The bottom panel shows the relative difference between the numerical solution and the experimental data.
}
\label{fig:ElectronPAMELAa}
\end{figure*}

\begin{figure*}[tb!]
 \includegraphics[width=0.49\textwidth]{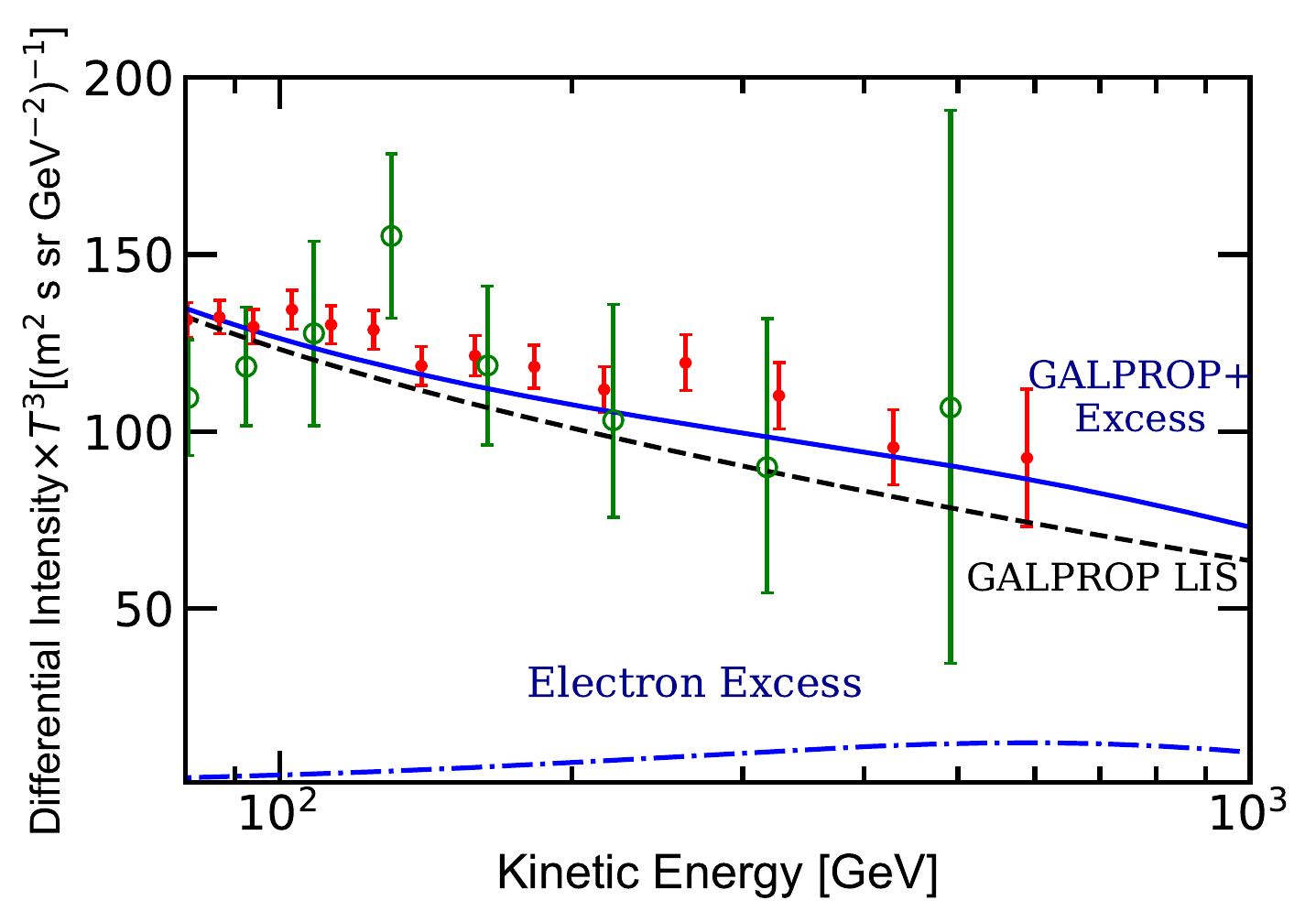}\hfill
 \includegraphics[width=0.49\textwidth]{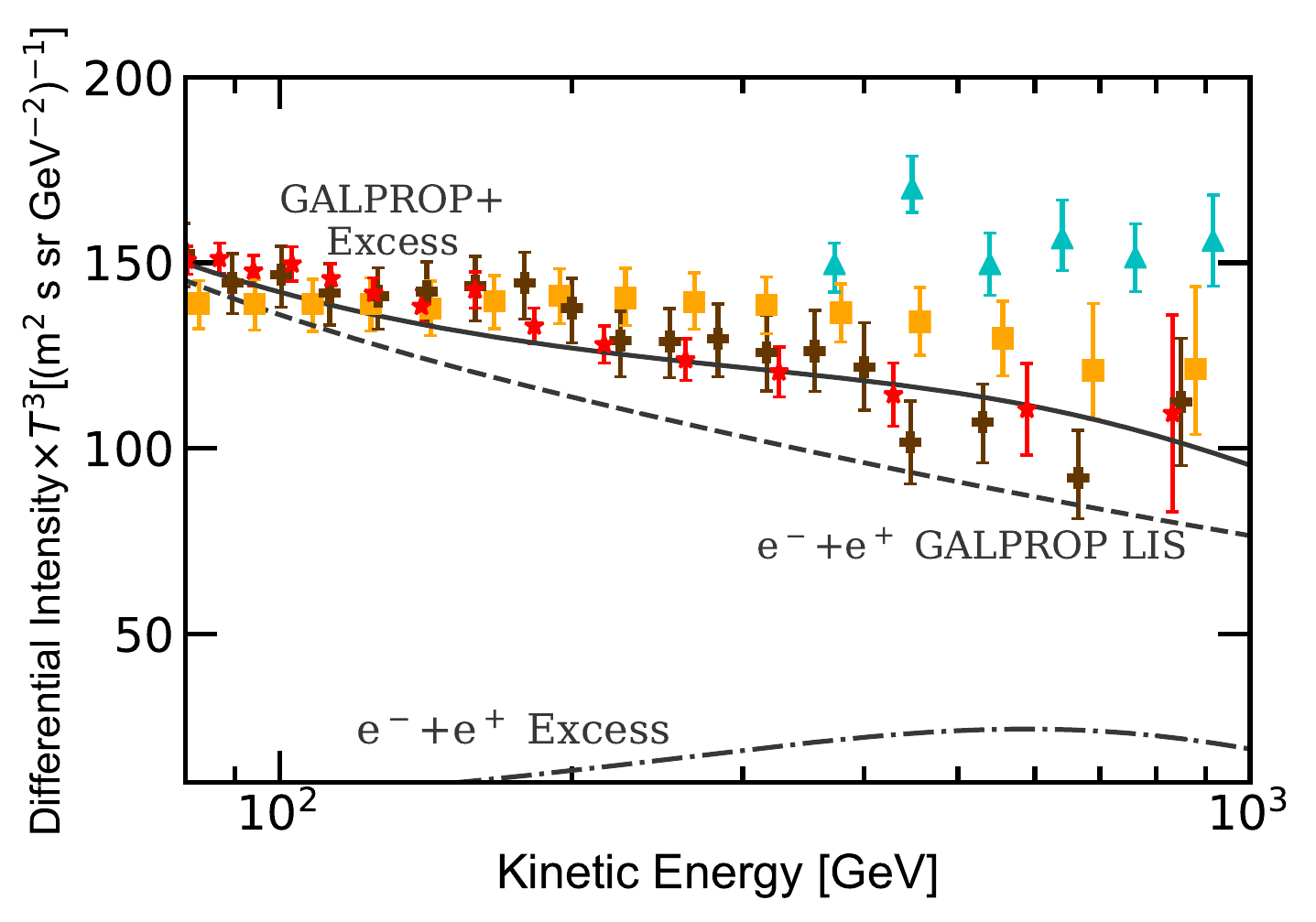}
\caption{High energy LIS for electrons (left panel) and ``all-electrons'' (right panel) along with AMS-02 (red points, \citealt{2014PhRvL.113l1102A,2014PhRvL.113v1102A}), PAMELA (green open circles, top panel,~\citealp{2011PhRvL.106t1101A}), \emph{Fermi}-LAT (orange squares, bottom panel, \citealt{2012PhRvL.108a1103A}), CALET (brown crosses, bottom panel, \citealt{PhysRevLett.119.181101}) and H.E.S.S.\ (cyan triangles, bottom panel, \citealt{2008PhRvL.101z1104A}) measurements. The \galprop{} LIS is plotted with the dashed line, the estimated high-energy omindirectional-intensity positron excess is plotted with the dot-dash line (multiplied by a factor of 2 in the bottom panel to account for the identical electron excess, see text) and, finally the sum of \galprop{} LIS and electron and positron excess components is plotted with the solid line.}
\label{fig:el_sig}
\end{figure*}

\subsection{Electron LIS at low and Intermediate energies}\label{Sect::LISOutsideMod}

Since the end of August, 2012, the Voyager 1 mission is exploring interstellar space providing invaluable data on the composition of Galactic CRs at low energies \citep{2013Sci...341..150S,2016ApJ...831...18C}. In the current analysis Voyager~1 data \citep{2016ApJ...831...18C} taken between December 2012 and June 2015 is used as a constraint for evaluating the electron LIS, as described above. A comparison of the Voyager 1 all-electron spectrum in the kinetic energy range 3--74 MeV and the proposed model for the LIS is shown in Figure~\ref{fig:GALPROPLISS}. The combined model provides a good description of the electron LIS at low energies. 

At high energies, where the CR fluxes are not affected by the heliospheric modulation, the most recent measurements by AMS-02 and PAMELA up to 90 GeV are included and shown in Figure~\ref{fig:low_high_LISData}. The electron LIS at even higher energies is discussed in Sect.~\ref{Sect::results}. 

It can be seen that even though the AMS-02 \citep{2014PhRvL.113l1102A} and PAMELA data \citep{2011PhRvL.106t1101A} in Figure~\ref{fig:low_high_LISData} are consistent within the error bars, the systematic difference between the datasets can be as large as $\sim$20\% in the energy range 30--90 GeV. Speculation on the possible origin(s) of this difference is not made here. However, it is clear that it is not the effect of solar modulation because it should be insignificant at these energies. For the MCMC procedure (Sect.~\ref{galprop}) the AMS-02 data is used because it has the smallest error bars.

\subsection{Data at Earth \mynote{and outside of the ecliptic plane}}\label{Sect::DataAtEarth}

%


%

This section illustrates an application of the HELMOD code to derivation of the modulated electron spectra at Earth. The spectra have to be compared to those measured by AMS-02 and PAMELA
during periods of low \citep[i.e., PAMELA from 2006 to 2010,][]{2011PhRvL.106t1101A,2015ApJ...810..142A} and high solar activity~\citep[i.e., AMS-02 from 2011 to 2013,][]{2014PhRvL.113l1102A}.
The available data are integrated over a period of a few months to years.
To reproduce the conditions of both low and high solar activity, the \helmod{} modulated spectra are evaluated for each Carrington Rotation within the period appropriate to the corresponding dataset.
The obtained results are then used to evaluate a unique normalized probability function for the modulation tool described in Section 3.1 of~\citet{2017ApJ...840..115B}.

Improvements in the data analysis procedure and in the simulation of the time dependence of the tracking system performance of PAMELA \citep{2015ApJ...810..142A} lead to a $\sim$10\% increase in the overall normalization of the CR electron fluxes measured in the period from July, 2006 -- December, 2009 compared to earlier results \citep{2011PhRvL.106t1101A}. However, it is not enough to account for a systematic discrepancy of $\sim$20\% between AMS-02 and earlier results from PAMELA \citep{2011PhRvL.106t1101A}. Due to the smaller quoted systematic uncertainties, the AMS-02 data are used as the reference. In this work a normalization factor for the electron LIS that is listed in Table~\ref{tbl-norm} is calculated for each presented dataset. 


The computed modulated spectra, for both low and high solar activity periods, are shown in Figures \ref{fig:ElectronAMSPAMELA} and \ref{fig:ElectronPAMELAa}. The details of the modulation model are described in Sect.~\ref{Sect::Helmod} and applied to the LIS described in Sect.~\ref{Sect::LISOutsideMod}. The high energy part of the spectrum is not affected by the solar modulation, and, therefore, is not discussed here. Simulated spectra are in a good agreement with experimental data in the energy range from 1 GeV to 90 GeV. The $\sim2\sigma$ deviations seen in the energy range $\la$3 GeV are present in all spectra, and this most likely implies that the injection spectrum needs some additional adjustments. Further comparison with the data is made in Appendix~\ref{app:SupMat} that also includes data taken by PAMELA around solar minimum \citep{2015ApJ...810..142A}.

\begin{figure*}[tb!]
\centerline{
 \includegraphics[width=0.95\textwidth]{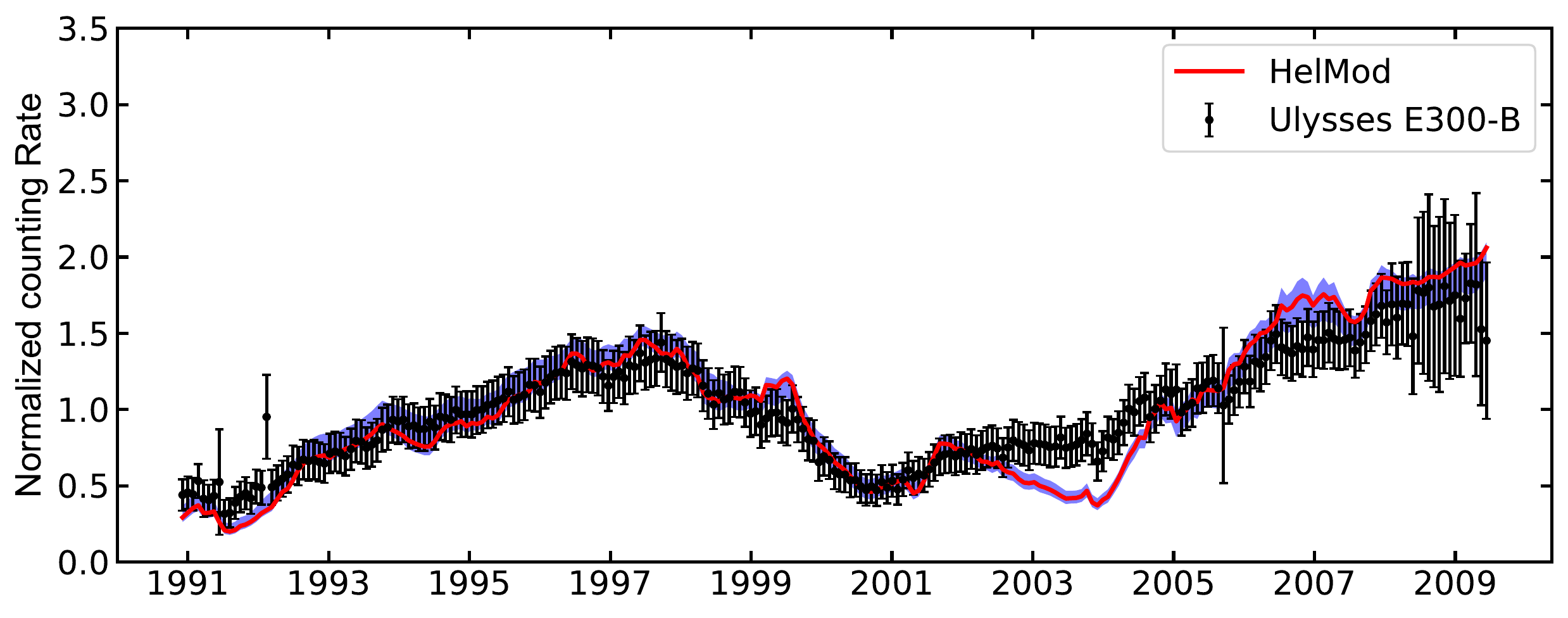}
 }
 \caption{\mynote{Ulysses counting rate normalized to the average value for the KET electron channel E300-B (electron energies of 0.9--4.6 GeV) as a function of time, where each point is an average over one Carrington rotation. The red solid line is the \helmod{} calculation for electrons of 0.6--10 GeV convolved with the subchannel response function for each Carrington rotation at same distance and solar latitude of the Ulysses spacecraft. Blue area shows systematic errors associated with model calculations. }}
 \label{fig:OutEcliptic}
\end{figure*}

\mynote{A reliable model for heliospheric modulation requires a proper modeling of CR distribution in the whole heliospheric volume, including space outside the ecliptic plane and at large distances from the Sun. Since 1990s and until 2009, the Ulysses spacecraft~\citep[see e.g.][]{Sandersonetal1995,Marsden2001,BaloghetAl2001} explored the heliosphere outside the ecliptic plane up to $\pm 80\degree$ in solar latitude and at distances $\sim$1--5 AU from the Sun. In particular, observations of particle flux were performed using the Cosmic Ray and Solar Particle Investigation Kiel Electron Telescope (COSPIN/KET) and High Energy Telescope (COSPIN/HET). Figure~\ref{fig:OutEcliptic} shows the Ulysses counting rate normalized to the average value. Data for Ulysses were taken from Ulysses Final Archive\footnote{http://ufa.esac.esa.int/ufa}. The analyzed data come from the KET electron channel E300-B \citep[electron energies of 0.9--4.6 GeV]{1996A&A...307..981R} using the Carrington Rotation average. }

\mynote{\helmod{} calculations are made for electrons of 0.6--10 GeV for each Carrington Rotation at the same distance and solar latitude as the Ulysses spacecraft. In order to correctly weight the spectral energy distribution, the calculated differential flux is then convolved with the subchannel response function available in \citet{1996A&A...307..981R}. The error band was evaluated using the procedure described in~\citet{2017ApJ...840..115B}. Figure~\ref{fig:OutEcliptic} shows a comparison of the Ulysses data with the \helmod{} calculations. Both experimental data and simulations are normalized to their corresponding mean values to allow a relative comparison along the solar cycle. The model reproduces the general features of the latitudinal gradients observed during the fast scans of 1994--1995 and 2007. Moreover, the agreement is still acceptable along the whole orbit, which extends as far as $\sim$3 AU. We note that the purpose of Figure~\ref{fig:OutEcliptic} is only to demonstrate the qualitative agreement between the \helmod{} calculations and observations. A proper quantitative comparison with the Ulysses data would require a calculation that combines several energy bins weighted with the Ulysses response function and detector efficiency.
}

\section{Electron LIS} \label{Sect::results}

In addition to the plots and tabulated data presented in Sect.~\ref{Sect::DataAtEarth} and the Appendix, we provide a parameterization $F(T)$ of the \galprop{} LIS (Figure \ref{fig:GALPROPLISS}) from 2 MeV up to 90 GeV as a function of kinetic energy in GeV:
\begin{eqnarray}\label{EQ::an}
&&F(T) = \\
&&\left\{ 
\begin{tabular}{lr}
	\multicolumn{2}{l}{$\displaystyle\frac{1.181\!\!\times\!\! 10^{11} T^{-12.061}}{1+4.307\!\!\times\!\! 10^{8} T^{-9.269}+3.125\!\!\times\!\! 10^{8}  T^{-10.697}}$,}\\
												& $T<6.88$ GeV\\
                $995.598\, T^{-3.505}+4.423\, T^{-2.620}$, & \quad\ $ T\ge 6.88$ GeV
                \end{tabular}
    \right.\nonumber
\end{eqnarray} 
%
where the units are (m$^2$ s sr GeV)$^{-1}$. This fit reproduces the \galprop{} electron LIS with an accuracy better than 5\% for the whole quoted energy range.

The electron LIS that results from the model calculations is in a good agreement with data (Figure \ref{fig:el_sig}). Meanwhile, it may harbor an additional electron component from an unknown source of the same nature as that of the excess positrons \citep{2009Natur.458..607A,2014PhRvL.113l1101A}. If charge-sign symmetry is assumed, i.e.\ that the electron and positron components coming from an unknown source have identical spectra, then the spectral shape of such an additional electron component can be derived from AMS-02 positron measurements \citep{2014PhRvL.113l1102A}. The spectrum of an additional component, ``the signal,'' $S(T)$ can be parametrized as a function of kinetic energy as:
\begin{equation}
S(T)= 4.5\times10^{-3}\ T^{-1.53} e^{-\frac{T}{400\ {\rm GeV}}}\ ({\rm m}^2\ {\rm s\ sr\ GeV})^{-1}
\end{equation}
This involves a re-tuning of the electron injection spectrum above the break at 95 GV ($\gamma_3$ in Table~\ref{tbl-2}). This parameterization also takes into account the 
standard astrophysical background of secondary positrons evaluated to be $\la$6\% at 30 GeV \citep{1998ApJ...493..694M,2014PhRvL.113l1101A}. 

With an addition of the extra components, the electron and all-electron spectra \citep{2014PhRvL.113v1102A} match the AMS-02 data well (Figure \ref{fig:el_sig}). The calculated all-electron spectrum includes the astrophysical background of positrons ($\la$6\% relative to the all-electron LIS) that is also used as an estimate of the  systematic uncertainty. The all-electron spectrum includes twice the positron excess that accounts for both extra electron and positron components. The inclusion of the extra electron and positron components in equal amounts improves the agreement with the AMS-02 data~\citep{RozzaJHEA_2015}. A possible origin of this excess will be discussed in a forthcoming paper devoted to the positron LIS.

\section{Conclusions}

The electron LIS derived in the current work provides a good description of the Voyager 1, PAMELA, and AMS-02 data over the energy range from 1 MeV to 1 TeV. The data for solar cycles 23 and 24 to be successfully reproduced within a single framework. This includes a fully realistic and exhaustive description of the relevant CR physics. Given their high precision, recent AMS-02 electron and positron data can be used to put useful constraints on the origin of the positron excess -- to be discussed in the forthcoming paper.  This work complements earlier results on the proton, He, and antiproton LIS illustrating a significant potential of the combined \galprop{}-\helmod{} framework.

\acknowledgements
Special thanks are made to Pavol Bobik, Giuliano Boella, Karel Kudela, Marian Putis and Mario Zannoni for their support of the \helmod{} code and many useful suggestions. This work is supported by ASI (Agenzia Spaziale Italiana) under contract ASI-INFN I/002/13/0 and ESA (European Space Agency) contract 4000116146/16/NL/HK. Igor Moskalenko and Troy Porter acknowledge support from NASA Grant No.~NNX17AB48G.

\bibliography{bibliography,imos}


\clearpage

\appendix  
\section{Supplementary Material} \label{app:SupMat}

\begin{figure*}[htb]
\centerline{
 \includegraphics[width=0.5\textwidth]{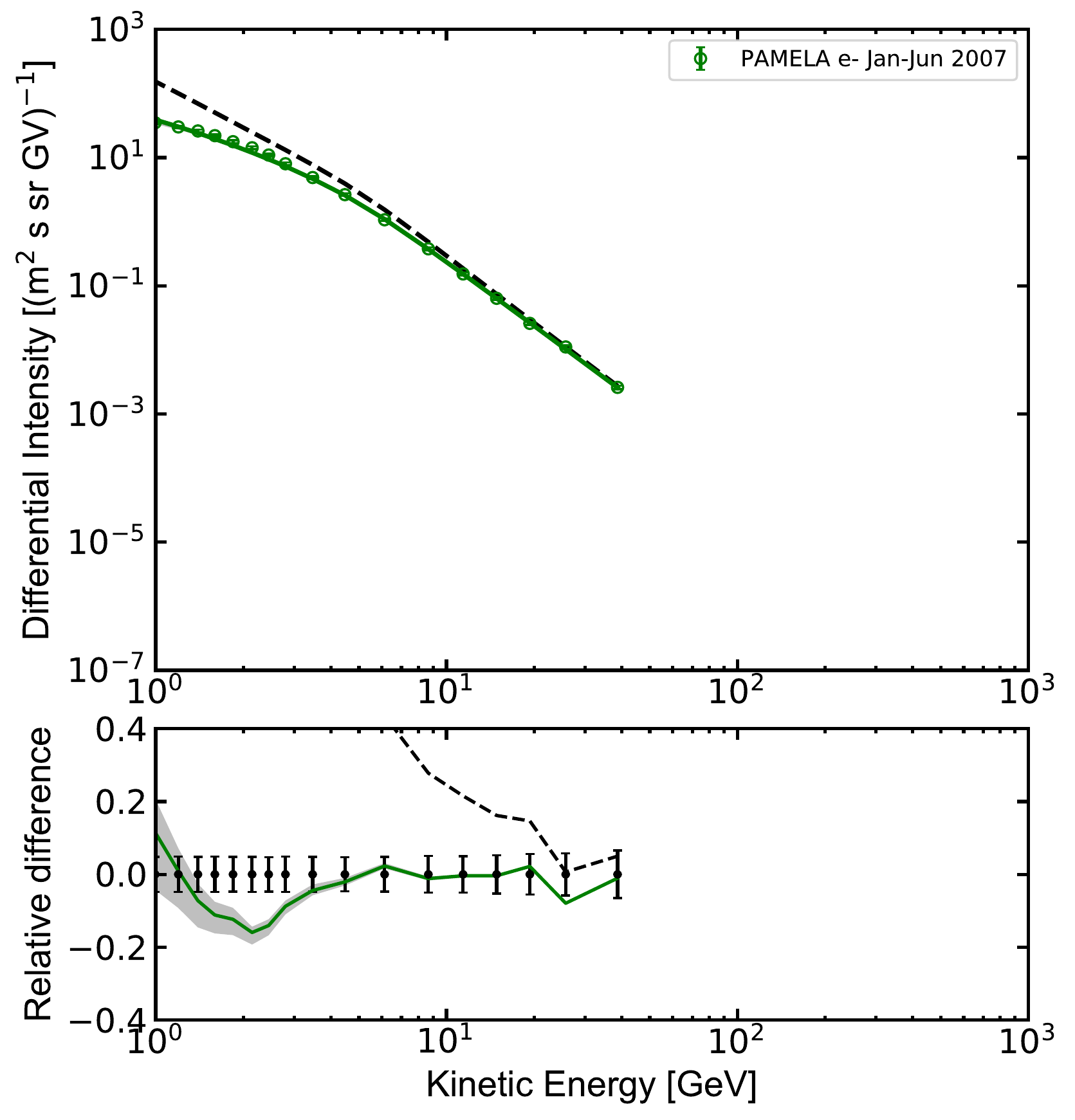}\hfill
 \includegraphics[width=0.5\textwidth]{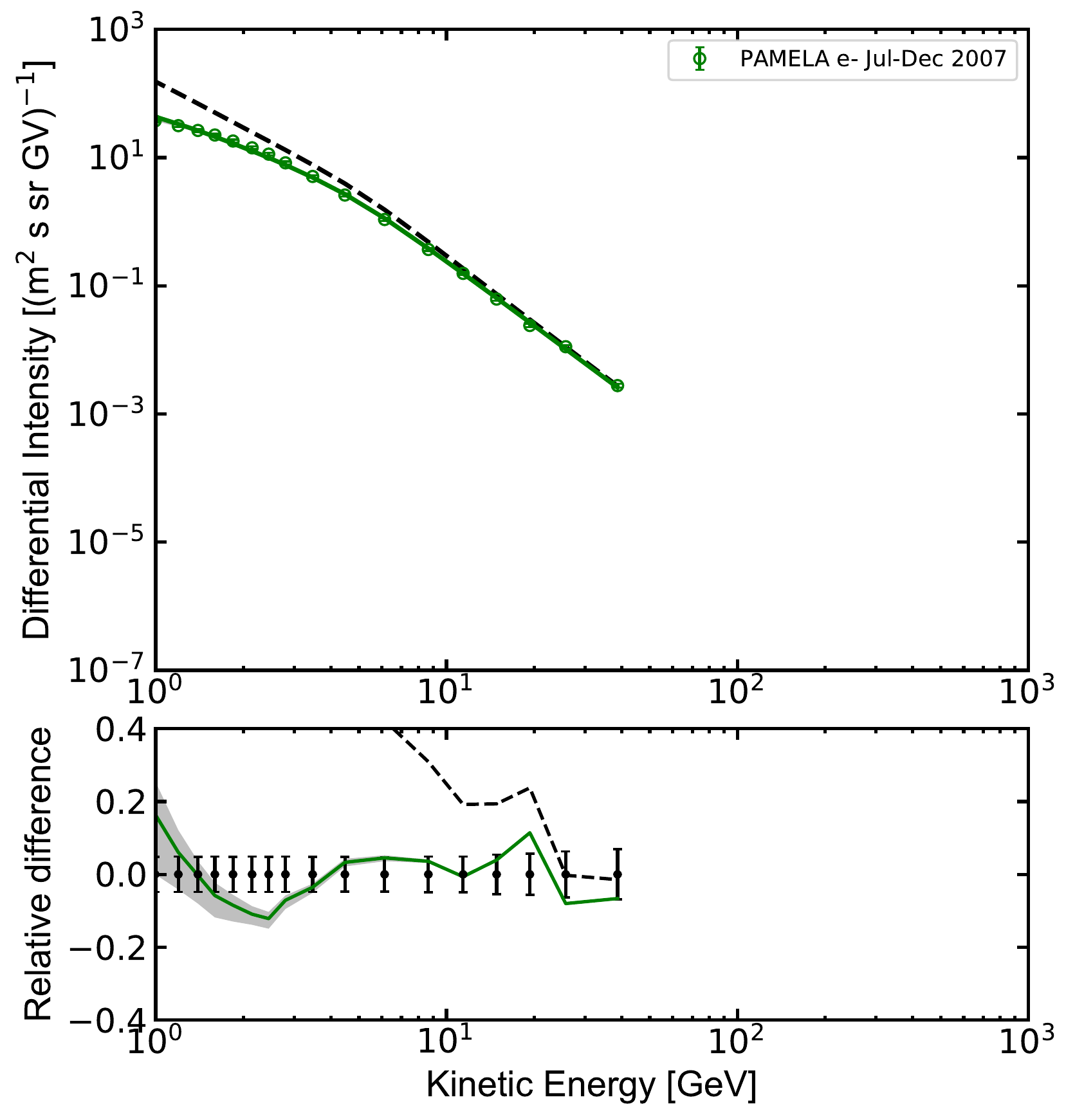}
 }
 \caption{Differential intensity of CR electrons for PAMELA 2007 datasets. Points represent experimental data, the black dashed line is the \galprop{} LIS, and the solid lines are the computed modulated spectra. The bottom panel shows the relative difference between the numerical solution and experimental data.
}
 \label{fig:ElectronPAMELAb}
\end{figure*}
\begin{figure*}[htb]
\centerline{
 \includegraphics[width=0.5\textwidth]{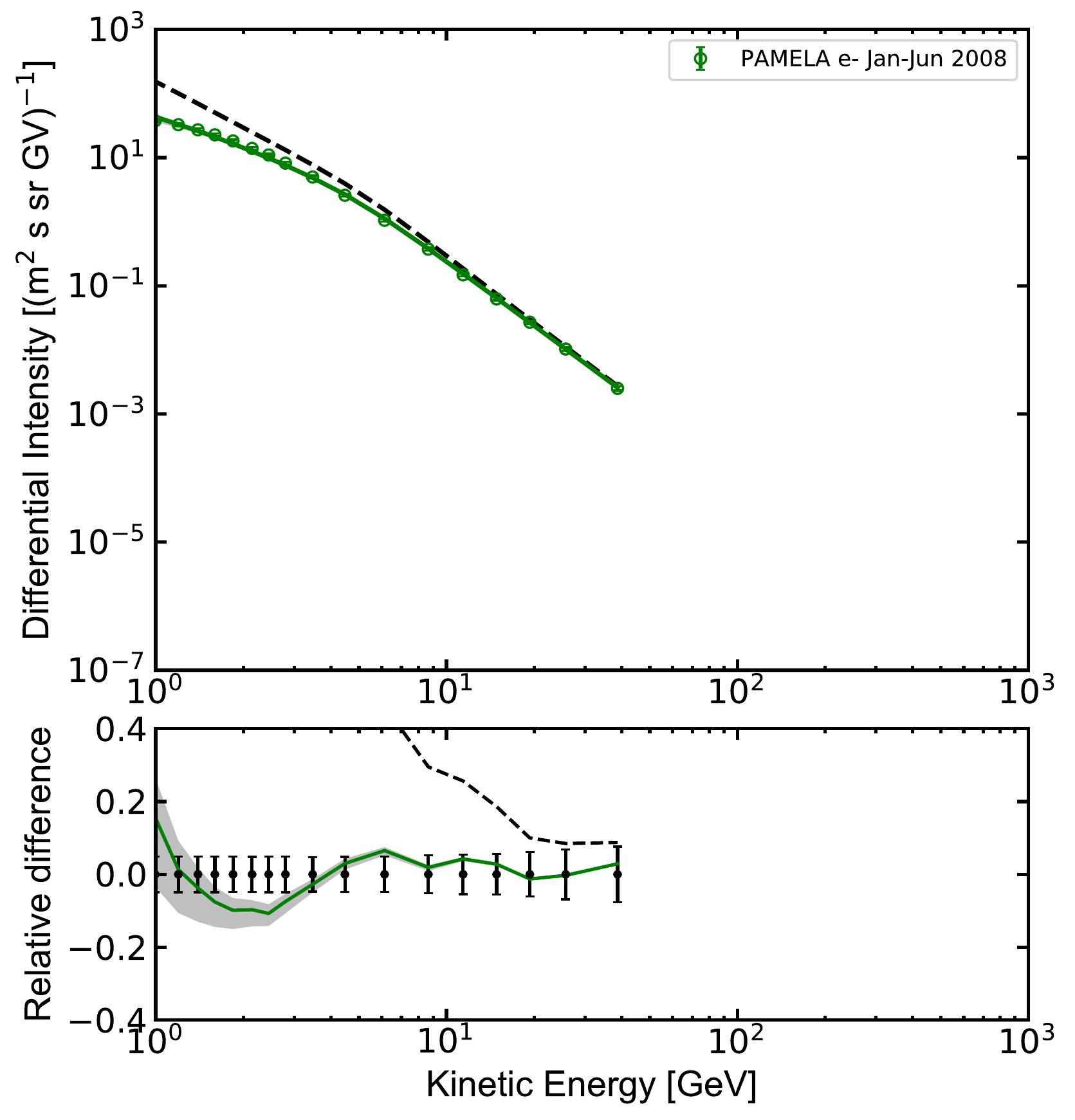}\hfill
 \includegraphics[width=0.5\textwidth]{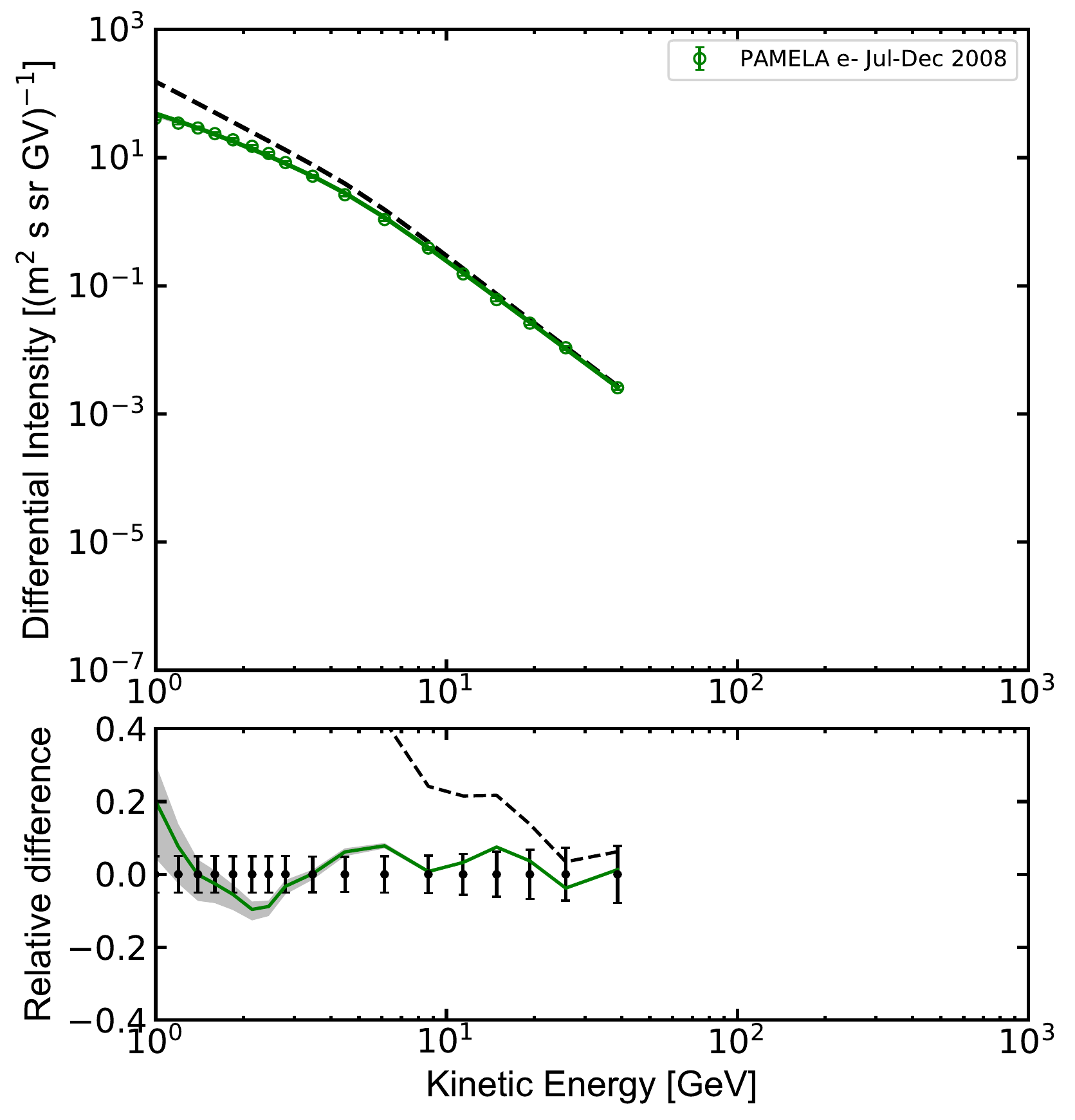}
 }
 \caption{Differential intensity of CR electrons for PAMELA 2008 datasets. Points represent experimental data, the dashed line is the \galprop{} LIS, and the solid line is the computed modulated spectrum. The bottom panel shows the relative difference between the numerical solution and experimental data.
}
 \label{fig:ElectronPAMELAc}
\end{figure*}

\begin{figure*}[htb]
\centerline{
 \includegraphics[width=0.5\textwidth]{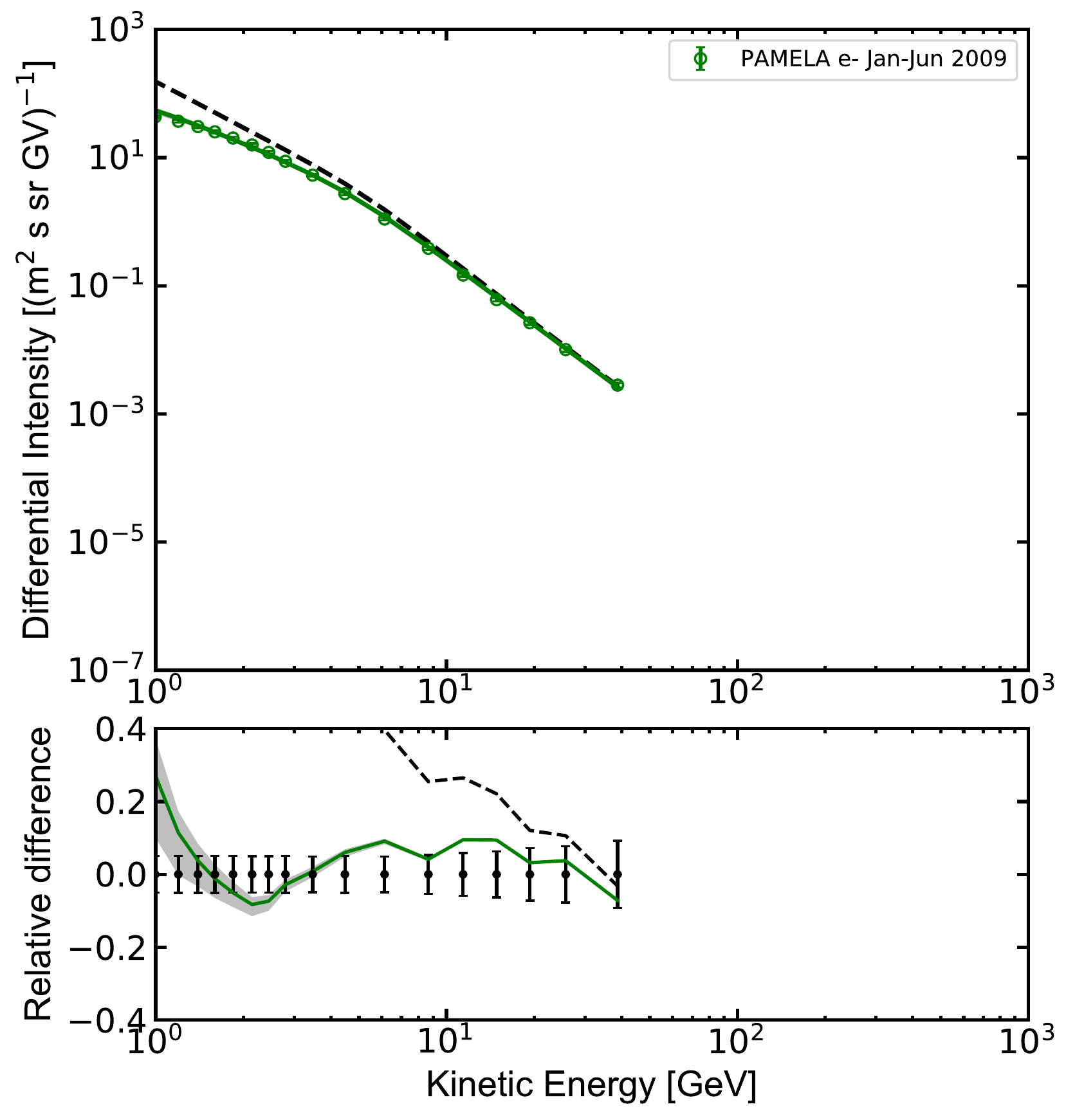}
 }
 \caption{Differential intensity of CR electrons for PAMELA datasets integrated from January to June 2009. Points represent experimental data, the black dashed line is the \galprop{} LIS, and the solid line is the computed modulated spectrum. The bottom panel shows the relative difference between the numerical solution and experimental data.
}
 \label{fig:ElectronPAMELAd}
\end{figure*}

\begin{deluxetable*}{ccccccccc}
\tablecolumns{10}
\tablewidth{0mm}
\tablecaption{Electron LIS\label{Tbl-ElectronLIS}}
\tablehead{
\colhead{Kinetic energy,} & \colhead{Differential} &
\colhead{Kinetic energy,} & \colhead{Differential} &
\colhead{Kinetic energy,} & \colhead{Differential} &
\colhead{Kinetic energy,} & \colhead{Differential} 
\\
\colhead{GeV} & \colhead{intensity\tablenotemark{a}} &
\colhead{GeV} & \colhead{intensity\tablenotemark{a}} &
\colhead{GeV} & \colhead{intensity\tablenotemark{a}} &
\colhead{GeV} & \colhead{intensity\tablenotemark{a}} 
}
\startdata
1.000e-03 & 3.481e+06 & 3.927e-02 & 3.225e+04 & 1.542e+00 & 6.057e+01 & 6.482e+01 & 5.236e-04 \\ 
1.070e-03 & 3.710e+06 & 4.203e-02 & 2.935e+04 & 1.651e+00 & 5.145e+01 & 6.938e+01 & 4.167e-04 \\ 
1.146e-03 & 3.523e+06 & 4.499e-02 & 2.671e+04 & 1.767e+00 & 4.370e+01 & 7.426e+01 & 3.318e-04 \\ 
1.226e-03 & 3.288e+06 & 4.815e-02 & 2.431e+04 & 1.891e+00 & 3.713e+01 & 7.949e+01 & 2.644e-04 \\ 
1.312e-03 & 3.056e+06 & 5.154e-02 & 2.212e+04 & 2.024e+00 & 3.155e+01 & 8.508e+01 & 2.108e-04 \\ 
1.405e-03 & 2.835e+06 & 5.517e-02 & 2.014e+04 & 2.166e+00 & 2.681e+01 & 9.106e+01 & 1.681e-04 \\ 
1.504e-03 & 2.627e+06 & 5.905e-02 & 1.834e+04 & 2.319e+00 & 2.278e+01 & 9.747e+01 & 1.342e-04 \\ 
1.609e-03 & 2.431e+06 & 6.320e-02 & 1.670e+04 & 2.482e+00 & 1.935e+01 & 1.043e+02 & 1.071e-04 \\ 
1.722e-03 & 2.246e+06 & 6.764e-02 & 1.521e+04 & 2.656e+00 & 1.642e+01 & 1.117e+02 & 8.560e-05 \\ 
1.844e-03 & 2.074e+06 & 7.240e-02 & 1.386e+04 & 2.843e+00 & 1.392e+01 & 1.195e+02 & 6.841e-05 \\ 
1.973e-03 & 1.912e+06 & 7.749e-02 & 1.262e+04 & 3.043e+00 & 1.179e+01 & 1.279e+02 & 5.470e-05 \\ 
2.112e-03 & 1.761e+06 & 8.295e-02 & 1.149e+04 & 3.257e+00 & 9.962e+00 & 1.369e+02 & 4.374e-05 \\ 
2.261e-03 & 1.621e+06 & 8.878e-02 & 1.046e+04 & 3.486e+00 & 8.397e+00 & 1.465e+02 & 3.499e-05 \\ 
2.420e-03 & 1.490e+06 & 9.502e-02 & 9.525e+03 & 3.732e+00 & 7.054e+00 & 1.569e+02 & 2.800e-05 \\ 
2.590e-03 & 1.368e+06 & 1.017e-01 & 8.669e+03 & 3.994e+00 & 5.904e+00 & 1.679e+02 & 2.240e-05 \\ 
2.772e-03 & 1.256e+06 & 1.089e-01 & 7.887e+03 & 4.275e+00 & 4.919e+00 & 1.797e+02 & 1.793e-05 \\ 
2.967e-03 & 1.151e+06 & 1.165e-01 & 7.173e+03 & 4.576e+00 & 4.076e+00 & 1.923e+02 & 1.435e-05 \\ 
3.176e-03 & 1.055e+06 & 1.247e-01 & 6.521e+03 & 4.898e+00 & 3.358e+00 & 2.059e+02 & 1.149e-05 \\ 
3.399e-03 & 9.658e+05 & 1.335e-01 & 5.926e+03 & 5.242e+00 & 2.748e+00 & 2.203e+02 & 9.193e-06 \\ 
3.638e-03 & 8.835e+05 & 1.429e-01 & 5.382e+03 & 5.611e+00 & 2.235e+00 & 2.358e+02 & 7.358e-06 \\ 
3.894e-03 & 8.077e+05 & 1.529e-01 & 4.886e+03 & 6.005e+00 & 1.806e+00 & 2.524e+02 & 5.889e-06 \\ 
4.168e-03 & 7.380e+05 & 1.637e-01 & 4.433e+03 & 6.428e+00 & 1.451e+00 & 2.702e+02 & 4.713e-06 \\ 
4.461e-03 & 6.739e+05 & 1.752e-01 & 4.019e+03 & 6.880e+00 & 1.160e+00 & 2.892e+02 & 3.772e-06 \\ 
4.775e-03 & 6.150e+05 & 1.875e-01 & 3.640e+03 & 7.364e+00 & 9.239e-01 & 3.095e+02 & 3.018e-06 \\ 
5.111e-03 & 5.609e+05 & 2.007e-01 & 3.295e+03 & 7.882e+00 & 7.335e-01 & 3.313e+02 & 2.415e-06 \\ 
5.470e-03 & 5.114e+05 & 2.148e-01 & 2.979e+03 & 8.436e+00 & 5.811e-01 & 3.546e+02 & 1.932e-06 \\ 
5.855e-03 & 4.660e+05 & 2.299e-01 & 2.690e+03 & 9.029e+00 & 4.598e-01 & 3.795e+02 & 1.545e-06 \\ 
6.267e-03 & 4.244e+05 & 2.461e-01 & 2.426e+03 & 9.665e+00 & 3.634e-01 & 4.062e+02 & 1.236e-06 \\ 
6.707e-03 & 3.864e+05 & 2.634e-01 & 2.185e+03 & 1.034e+01 & 2.871e-01 & 4.348e+02 & 9.887e-07 \\ 
7.179e-03 & 3.517e+05 & 2.819e-01 & 1.964e+03 & 1.107e+01 & 2.268e-01 & 4.654e+02 & 7.907e-07 \\ 
7.684e-03 & 3.200e+05 & 3.018e-01 & 1.762e+03 & 1.185e+01 & 1.791e-01 & 4.981e+02 & 6.323e-07 \\ 
8.225e-03 & 2.910e+05 & 3.230e-01 & 1.577e+03 & 1.268e+01 & 1.415e-01 & 5.332e+02 & 5.056e-07 \\ 
8.803e-03 & 2.646e+05 & 3.457e-01 & 1.409e+03 & 1.358e+01 & 1.118e-01 & 5.707e+02 & 4.042e-07 \\ 
9.422e-03 & 2.406e+05 & 3.700e-01 & 1.255e+03 & 1.453e+01 & 8.829e-02 & 6.108e+02 & 3.231e-07 \\ 
1.008e-02 & 2.186e+05 & 3.960e-01 & 1.115e+03 & 1.555e+01 & 6.976e-02 & 6.538e+02 & 2.583e-07 \\ 
1.079e-02 & 1.987e+05 & 4.239e-01 & 9.886e+02 & 1.665e+01 & 5.512e-02 & 6.997e+02 & 2.064e-07 \\ 
1.155e-02 & 1.805e+05 & 4.537e-01 & 8.739e+02 & 1.782e+01 & 4.356e-02 & 7.489e+02 & 1.650e-07 \\ 
1.237e-02 & 1.640e+05 & 4.856e-01 & 7.703e+02 & 1.907e+01 & 3.444e-02 & 8.016e+02 & 1.318e-07 \\ 
1.324e-02 & 1.489e+05 & 5.198e-01 & 6.771e+02 & 2.041e+01 & 2.722e-02 & 8.580e+02 & 1.053e-07 \\ 
1.417e-02 & 1.353e+05 & 5.563e-01 & 5.934e+02 & 2.185e+01 & 2.152e-02 & 9.184e+02 & 8.416e-08 \\ 
1.516e-02 & 1.229e+05 & 5.955e-01 & 5.185e+02 & 2.339e+01 & 1.703e-02 & 9.830e+02 & 6.724e-08 \\ 
1.623e-02 & 1.116e+05 & 6.374e-01 & 4.517e+02 & 2.503e+01 & 1.347e-02 & 1.052e+03 & 5.372e-08 \\ 
1.737e-02 & 1.014e+05 & 6.822e-01 & 3.923e+02 & 2.679e+01 & 1.066e-02 & 1.126e+03 & 4.291e-08 \\ 
1.859e-02 & 9.206e+04 & 7.302e-01 & 3.398e+02 & 2.867e+01 & 8.439e-03 & 1.205e+03 & 3.427e-08 \\ 
1.990e-02 & 8.363e+04 & 7.815e-01 & 2.934e+02 & 3.069e+01 & 6.680e-03 & 1.290e+03 & 2.737e-08 \\ 
2.130e-02 & 7.597e+04 & 8.365e-01 & 2.527e+02 & 3.285e+01 & 5.304e-03 & 1.381e+03 & 2.186e-08 \\ 
2.280e-02 & 6.903e+04 & 8.953e-01 & 2.171e+02 & 3.516e+01 & 4.192e-03 & 1.478e+03 & 1.746e-08 \\ 
2.440e-02 & 6.273e+04 & 9.583e-01 & 1.861e+02 & 3.763e+01 & 3.316e-03 & 1.582e+03 & 1.394e-08 \\ 
2.612e-02 & 5.701e+04 & 1.026e+00 & 1.592e+02 & 4.028e+01 & 2.633e-03 & 1.693e+03 & 1.113e-08 \\ 
2.796e-02 & 5.182e+04 & 1.098e+00 & 1.359e+02 & 4.311e+01 & 2.077e-03 & 1.812e+03 & 8.888e-09 \\ 
2.992e-02 & 4.712e+04 & 1.175e+00 & 1.158e+02 & 4.615e+01 & 1.649e-03 & 1.940e+03 & 7.096e-09 \\ 
3.203e-02 & 4.285e+04 & 1.258e+00 & 9.861e+01 & 4.939e+01 & 1.310e-03 & 2.076e+03 & 5.665e-09 \\ 
3.428e-02 & 3.897e+04 & 1.346e+00 & 8.387e+01 & 5.287e+01 & 1.042e-03 & 2.222e+03 & 4.522e-09 \\ 
3.669e-02 & 3.545e+04 & 1.441e+00 & 7.129e+01 & 5.658e+01 & 8.282e-04 & 2.378e+03 & 3.610e-09  
\enddata
\tablenotetext{a}{Differential intensity units: (m$^2$ s sr GV)$^{-1}$.}
\end{deluxetable*}


\end{document}